\documentclass[pdflatex,sn-basic]{sn-jnl}

\jyear{2021}%

\theoremstyle{thmstyleone}%
\theoremstyle{thmstyletwo}%

\theoremstyle{thmstylethree}%

\newcommand{\argmin}{\mathop{\rm arg~min}\limits}
\providecommand{\abs}[1]{\lvert#1\rvert}

\usepackage{bm}
\usepackage{lscape}
\usepackage{etoolbox}
\usepackage[subrefformat=parens]{subcaption}

\makeatletter
\patchcmd{\ps@headings}
{\hbox to \hsize{\hfill Springer Nature 2021 \LaTeX\ template\hfill}}
{\hbox to \hsize{\hfill \hfill}}
{}
{}
\makeatother
\makeatletter
\patchcmd{\ps@headings}
{\hbox to \hsize{\hfill Springer Nature 2021 \LaTeX\ template\hfill}}
{\hbox to \hsize{\hfill \hfill}}
{}
{}
\makeatother
\makeatletter
\patchcmd{\ps@headings}
{\hbox to \hsize{\hfill Springer Nature 2021 \LaTeX\ template\hfill}}
{\hbox to \hsize{\hfill \hfill}}
{}
{}
\makeatother

\begin{document}

\title[]{Multi-task Learning for Compositional Data via Sparse Network Lasso}


\author*[1]{\fnm{Akira} \sur{Okazaki}}\email{okazaki.akira@ai.lab.uec.ac.jp}

\author[1]{\fnm{Shuichi} \sur{Kawano}}

\affil[1]{\orgdiv{Graduate School of Informatics and Engineering}, \orgname{The University of Electro-Communications}, \orgaddress{\street{1-5-1 Chofugaoka}, \city{Chofu}, \state{Tokyo}, \postcode{182-8585} \country{Japan}}}


\abstract{
A network lasso enables us to construct a model for each sample, which is known as multi-task learning. Existing methods for multi-task learning cannot be applied to compositional data due to their intrinsic properties. 
In this paper, we propose a multi-task learning method for compositional data using a sparse network lasso. We focus on a symmetric form of the log-contrast model, which is a regression model with compositional covariates. The effectiveness of the proposed method is shown through simulation studies and application to gut microbiome data.}

\keywords{Clustering, Log-contrast model, Multi-task learning, Symmetric form, Regularization, Variable selection.}



\maketitle

\section{Introduction}\label{sec1}

Multi-task learning is a methodology that assumes a different model for each task and estimates these models. It is used in various fields of research, in particular life science (\cite{Argyriou2008-cj}; \cite{Lengerich2018-zo}). For example, the factors of a disease may vary from patient to patient \citep{Cowie1997-ib}. In such a case, a model that is common to all patients cannot sufficiently extract the factors of the disease. In multi-task learning, each patient is considered as a single task, and different models can be built for each patient to extract both patient-specific and common factors for the disease \citep{Xu2015-yb}. \cite{Yamada2017-cg} proposed the localized lasso, which performs multi-task learning using network lasso regularization \citep{Hallac2015-ss}. By treating each sample as a single task, localized lasso simultaneously performs multi-task learning and clustering in the framework of a regression model.

 On the other hand, compositional data, which consist of the proportions of a composition, are used in the fields of geology and life science for microbiome analysis. Compositional data are constrained to always take positive values summing to one. Due to these constraints, it is difficult to apply existing multi-task learning methods to compositional data. In the field of microbiome analysis, studies on gut microbiomes (\cite{Wu2011-os}; \cite{Dillon2016-ph}) have suggested that there are multiple types of gut microbiome clusters that vary from individual to individual \citep{Arumugam2011-si}. In the case of such data where multiple clusters may exist, it is difficult to extract sufficient information using existing regression models for compositional data.

In this paper, we propose a multi-task learning method for compositional data, focusing on network lasso regularization and the symmetric form of the log-contrast model \citep{Aitchison1984-tl}, which is a linear regression model with compositional covariates. The symmetric form is extended to the locally symmetric form in which each sample has a different regression coefficient vector. These regression coefficient vectors are clustered by network lasso regularization. In addition, because dimensionality of features of compositional data has been increasing \citep{Lin2014-ii}, we use $\ell_{1}$ regularization \citep{Tibshirani1996-ak} to perform variable selection. Estimation of the parameters included in the model is performed using an estimation algorithm based on the alternating direction method of multipliers \citep{Boyd2011-ul}, because the model includes non-differentiable points in the $\ell_{1}$ regularization term and zero-sum constraints on the parameters. The constructed model includes regularization parameters, which are determined by cross-validation (CV).

The remainder of this paper is organized as follows. Section 2 introduces multi-task learning based on a network lasso. In Section 3, we describe the regression models for compositional data. We propose a multi-task learning method for compositional data and its estimation algorithm in Section 4. In Section 5, we discuss the effectiveness of the proposed method through Monte Carlo simulations. An application to gut microbiome data is presented in Section 6. Finally, Section 7 summarizes this paper and discusses future work.

\section{Multi-task learning based on a network lasso}\label{sec2}

Suppose that we have $n$ observed $p$-dimensional data $\{\bm{x}_{i};i=1,\ldots,n\}$ and $n$ observed data for the response variable $\{y_{i};i=1,\ldots,n\}$, and these pairs $\{({y}_{i},\bm{x}_{i}),i=1,\ldots,n\}$ are given independently. The graph $R=R^{T}\in\mathbb{R}^{n\times n}$ is also given, where $(R)_{ij}=r_{i,j}\geq 0$ represents the relationship between the sample pair $(y_{i},\bm{x}_{i})$ and $(y_{j},\bm{x}_{j})$, and thus the diagonal components are zero.

We consider the linear regression model 
 \begin{equation}
    \label{localreg}
     y_{i} = \bm{x}_{i}^{T}\bm{w}_{i} + \epsilon_{i},\quad i=1,\ldots,n,
 \end{equation}
 where $\bm{w}_{i}=(w_{i1},\ldots,w_{ip})\in\mathbb{R}^{p}$ is the $p$-dimensional regression coefficient vector for sample $\bm{x}_{i}$, and $\epsilon_{i}$ is an error term distributed as $N(0,\sigma^{2})$ independently. Model (\ref{localreg}) comprises a different model for each sample. In general, the regression coefficient vectors are assumed to be the identical (i.e., $\bm{w}_{1}=\bm{w}_{2}=\cdots=\bm{w}_{n}$). For Model (\ref{localreg}), we consider the minimization problem
 \begin{equation}
    \label{netlasso}
     \min_{\bm{w}_{i}} \left\{ \sum_{i=1}^{n} (y_{i}-\bm{x}_{i}^{T}\bm{w}_{i})^{2}+\lambda  \sum_{i\neq j}^{n}r_{i,j} \|\bm{w}_{i}-\bm{w}_{j} \|_{2} \right\},
 \end{equation}
 where $\lambda \ (> 0)$ is a regularization parameter. 
 The second term in \eqref{netlasso} is the network lasso regularization term \citep{Hallac2015-ss}. 
 For coefficient vectors $\bm{w}_{i}$ and $\bm{w}_{j}$, the network lasso regularization term induces $\bm{w}_{i}=\bm{w}_{j}$. 
 If these vectors are estimated to be the same, then the $i$-th and $j$-th samples are interpreted as belonging to the same cluster. In the framework of multi-task learning, the minimization problem (\ref{netlasso}) considers one sample as one task by setting a coefficient vector for each sample. This allows us to extract the information of the regression coefficient vectors separately for each task. In addition, by clustering the regression coefficient vectors using the network lasso regularization term, we can extract the common information among tasks.
 
 \cite{Yamada2017-cg} proposed the localized lasso for minimization problem (\ref{netlasso}) by adding an $\ell_{1,2}$-norm regularization term \citep{Kong2014-jm}, which induces sparsity and group structure. The localized lasso is used for multi-task learning and variable selection.

\section{Regression modeling for compositional data}\label{sec3}
The $p$-dimensional compositional data $\bm{x}=(x_{1},\ldots,x_{p})^{T}$ are defined as proportional data in the simplex space
  \begin{equation}
     \mathbb{S}^{p-1} = \left\{ (x_{1},\ldots,x_{p}):x_{j}>0\quad (j=1,\ldots,p),\sum_{j=1}^{p}x_{j}=1 \right\}.
    \end{equation}
  This structure imposes dependence between the features of the compositional data. Thus, statistical methods defined for spaces of real numbers cannot be applied \citep{Aitchison1982-nl}. To overcome this problem, \citet{Aitchison1984-tl} proposed the log-contrast model, which is a linear regression model with compositional covariates.
    
    Suppose that we have $n$ observed $p$-dimensional compositional data $\{\bm{x}_{i};i=1,\ldots,n\}$ and $n$ objective variable data $\{y_{i};i=1,\ldots,n\}$, and these pairs $\{({y}_{i},\bm{x}_{i}),i=1 \ldots,n\}$ are given independently. The log-contrast model is represented as follows:
      \begin{equation}
      \label{log-cont}
        y_{i} = \sum_{i=1}^{p-1}\log\frac{x_{ij}}{x_{ip}}\beta_{i} +\epsilon_{i}, \quad i=1,\ldots,n,
      \end{equation}
     where $\bm{\beta} =(\beta_{1},\ldots,\beta_{p-1})^{T}\in\mathbb{R}^{p-1}$ is a regression coefficient vector. Because the model uses an arbitrary variable as a reference for all other variables, the solution changes depending on the selection of the reference. By introducing $\beta_{p}=-\sum_{j=1}^{p}\beta_{j}$, the log-contrast model is equivalently expressed in symmetric form as
    \begin{equation}
    \label{sym}
        y_{i} = \bm{z}_{i}^{T}\bm{\beta} +\epsilon_{i} ,\quad {\mathrm{s.t.}}\quad\sum_{j=1}^{p}\beta_{j}=0, \quad i=1,\ldots,n,
    \end{equation}
    where $\bm{z}_{i}=(\log x_{i1},\ldots,\log x_{ip})^{T}$, and $\bm{\beta}=(\beta_{1},\ldots,\beta_{p})^{T}\in\mathbb{R}^{p}$ is a regression coefficient vector. \citet{Lin2014-ii} proposed the minimization problem to select relevant variables in symmetric form by adding an $\ell_{1}$ regularization term \citep{Tibshirani1996-ak}:
    \begin{equation}
        \label{syml1}
        \min_{\bm{\beta}\in\mathbb{R}^{p}} \left\{ \sum_{i=1}^{n}
        (y_{i} - \bm{z}_{i}^{T}\bm{\beta})^{2}+\lambda\|\bm{\beta}\|_{1}\right\},\quad {\mathrm{s.t.}}\quad\sum_{j=1}^{p}\beta_{j}=0.
    \end{equation}
    Other models that extend this symmetric form of the problem have also been proposed (\cite{Shi2016-ys}; \cite{Wang2017-sy}; \cite{Bien2021-cl}; \cite{Combettes2021-br}).

\section{Proposed method}\label{sec4}

In this section, we propose a multi-task learning method for compositional data based on the network lasso and the symmetric form of the log-contrast model.

\subsection{Model}

We consider the locally symmetric form of the log-contrast model
    \begin{equation}
    \label{losym}
            y_{i} =  \bm{z}_{i}^{T}\bm{w}_{i}+\epsilon_{i},\quad {\mathrm{s.t.}}\quad \sum_{j=1}^{p}w_{ij}=0, \quad \quad i=1,\ldots,n,
    \end{equation}
where $\bm{w}_{i}=(w_{i1},\ldots,w_{ip})^{T}$ is the regression coefficient vector for $i$-th sample $\bm{x}_{i}$. 
For Model (\ref{losym}), we consider the minimization problem
    \begin{equation}
        \begin{split}
        \label{prop}
            \min_{\bm{w}_{i}\in \mathbb{R}^{p},\:i=1\ldots,n}
            \left\{ \sum_{i=1}^{n} (y_{i} - \bm{z}_{i}^{T}\bm{w}_{i} )^{2} +\lambda_{1}\sum_{i,j=1}^{n}r_{i,j}\|\bm{w}_{i}-\bm{w}_{j} \|_{2}+\lambda_{2}\sum_{i=1}^{n}\| \bm{w}_{i}\|_{1} \right\}, \\
            {\mathrm{s.t.}} \quad \sum_{j=1}^{p}w_{ij}=0,\quad i=1,\ldots,n, 
        \end{split}
    \end{equation}
    where $\lambda_{1},\lambda_{2}\:(>0)$ are regularization parameters. The second term is the network lasso regularization term, which performs the clustering of the regression coefficient vectors. The third term is the $\ell_{1}$ regularization term, which performs variable selection.
    
    A regression coefficient vector $\bm{w}_{i^{\ast}}=(\bm{w}_{i^{\ast}1},\ldots,\bm{w}_{i^{\ast}p})^{T}$ for future data $\bm{z}_{i^{\ast}}$ is obtained from the solution to the constrained Weber problem
    \begin{equation}
        \label{c-weber}
        \min_{\bm{w}_{i^{\ast}}\in \mathbb{R}^{p}}  \left\{ \sum_{i=1}^{n} r_{i^{\ast},i}  \| \bm{w}_{i^{\ast}} - \widehat{\bm{w}}_{i} \|_{2}, \right\}\quad{\mathrm{s.t.}}\sum_{j=1}^{p}w_{i^{\ast}j}=0,
    \end{equation}
    where $\widehat{\bm{w}}_{i}$ is the estimated regression coefficient vector for the $i$-th sample.
    \subsection{Estimation algorithm}
    To construct the estimation algorithm for the proposed method, we rewrite minimization problem (\ref{prop}) as follows:
    \begin{equation}
        \begin{split}
            \label{min2}
            \min_{\bm{w}_{i},\bm{a}_{i},\bm{b}_{i}\in \mathbb{R}^{p},\:i=1\ldots,n}
            \left\{ \sum_{i=1}^{n} (y_{i} - \bm{z}_{i}^{T}\bm{w}_{i} )^{2} +\lambda_{1}\sum_{i,j=1}^{n}r_{i,j}\|\bm{a}_{i,j}-\bm{a}_{j,i} \|_{2}+\lambda_{2}\sum_{i=1}^{n}\| \bm{b}_{i}\|_{1} \right\}, \\
            {\mathrm{s.t.}} \quad \bm{w}_{i} = \bm{a}_{i,j},\quad\bm{w}_{i} = \bm{b}_{i},\quad \bm{1}^{T}_{p}\bm{w}_{i}=0,\quad i=1,\ldots,n,
        \end{split}
    \end{equation}
    where $\bm{1}_{p}$ is the $p$-dimensional vector of ones. The augmented Lagrangian for (\ref{min2}) is formulated as
    \begin{equation}
        \label{lagrange}
        \begin{split}
            L(\Theta,Q)_{\Omega} &= \sum_{i=1}^{n} (y_{i} - \bm{w}_{i}^{T}\bm{z}_{i})^{2} \\ 
            &+\sum_{i>j}^{n} \left\{\lambda_{1} r_{i,j}\|\bm{a}_{i,j}-\bm{a}_{j,i} \|_{2}\right. \\
            &+\left.\frac{\rho}{2}(\|\bm{w}_{i}-\bm{a}_{i,j}+\bm{s}_{i,j} \|_{2}^{2}+\|\bm{w}_{j}-\bm{a}_{j,i}+\bm{s}_{j,i}\|_{2}^{2})-\frac{\rho}{2}(\|\bm{s}_{i,j}\|_{2}^{2}+\|\bm{s}_{j,i}\|_{2}^{2}) \right\} \\
            &+ \sum_{i=1}^{n} \left\{ \lambda_{2}\|\bm{b}_{i} \|_{1}+\bm{t}_{i}^{T}(\bm{w}_{i}-\bm{b}_{i})+\frac{\phi}{2}\|\bm{w}_{i}-\bm{b}_{i} \|_{2}^{2} \right\} \\
            &+\sum_{i=1}^{n}\left\{ u_{i}\bm{1}_{p}^{T}\bm{w}_{i} + \frac{\psi}{2}\|\bm{1}_{p}^{T}\bm{w}_{i}\|_{2}^{2}\right\},
        \end{split}
    \end{equation}
  where $\bm{s}_{i,j},\bm{t}_{i},u_{i}$ are the Lagrange multipliers and $\rho,\phi,\psi\:(>0)$ are the tuning parameters. For simplicity of notation, the parameters in the model $\bm{w}_{i},\bm{a}_{i,j},\bm{b}_{i}$ are collectively denoted as $\Theta$, the Lagrange multipliers as $Q$, and the tuning parameters as $\Omega$.
  
The update algorithm for $\Theta$ with the alternating direction method of multipliers (ADMM) is obtained from the minimization problem

    \begin{equation}
        \label{arglagsplit}
        \begin{split}
            \bm{w}^{(k+1)} &= \argmin_{\bm{w}}L(\bm{w},\bm{a}^{(k)},\bm{b}^{(k)},Q^{(k)})_{\Omega} \\
            \bm{a}^{(k+1)} &= \argmin_{\bm{a}}L(\bm{w}^{(k+1)},\bm{a},\bm{b}^{(k)},Q^{(k)})_{\Omega}, \\
             \bm{b}^{(k+1)} &= \argmin_{\bm{b}}L(\bm{w}^{(k+1)},\bm{a}^{(k+1)},\bm{b},Q^{(k)})_{\Omega},
        \end{split}
    \end{equation}
  where $k$ denotes the repetition number. By repeating the updates \eqref{arglagsplit} and the update for $Q$, the estimation algorithm for \eqref{min2} is represented by Algorithm 1. The estimation algorithm for \eqref{c-weber} is represented by Algorithm 2 with the update from ADMM. 
  The details of the derivations of the estimation algorithms are presented in Appendices A and B.
\begin{algorithm}[H]
    \caption{Estimation algorithm for (\ref{min2}) via ADMM}
    \label{Proposedalgo}
    \begin{algorithmic}
      \Require Initialize\:$\bm{w}^{(0)},\bm{a}^{(0)},\bm{b}^{(0)},\bm{s}^{(0)},\bm{t}^{(0)},u^{(0)}$.
      \While{convergence}
        \For{$i=1\ldots,n$}
        \State \begin{equation*}
            \begin{split}
                \bm{w}_{i}^{(k+1)}&=\left\{ 2\bm{z}_{i}\bm{z}_{i}^{T}+(\rho(n-1)+\phi)I_{p} +  \psi\bm{1}_{p}\bm{1}_{p}^{T}\right\}^{-1}\\&\left\{2 y_{i}\bm{z}_{i}+\rho \sum_{i>j}^{n}(\bm{a}_{i,j}^{(k)}-\bm{s}_{j,i}^{(k)})-\bm{t}_{i}^{(k)}+\phi\bm{b}_{i}^{(k)}-u_{i}\bm{1}_{p} \right\}
            \end{split}
        \end{equation*}
        \EndFor
        \For{$i,j=1,\ldots,n,\:(i>j)$}
        \State $ \theta = \mathrm{max}\left(1-\frac{\lambda_{1}r_{i,j}}{\rho\|(\bm{w}_{i}^{(k+1)}+\bm{s}_{i,j}^{(k)})-(\bm{w}_{j}^{(k+1)}+\bm{s}_{j,i}^{(k)}) \|_{2}},0.5 \right) $
        \State $\bm{a}_{i,j}^{(k+1)} = \theta (\bm{w}_{i}^{(k+1)}+\bm{s}_{i,j}^{(k)})+(1-\theta)(\bm{w}_{j}^{(k+1)}+\bm{s}_{j,i}^{(k)}) $
        \State $\bm{a}_{j,i}^{(k+1)} = (1-\theta) (\bm{w}_{i}^{(k+1)}+\bm{s}_{i,j}^{(k)})+\theta(\bm{w}_{j}^{(k+1)}+\bm{s}_{j,i}^{(k)}) $
        \EndFor
        \For{$i=1,\ldots,n,\:j=1,\ldots,p$}
          \State $b_{ij}^{(k+1)} = \mathrm{S}(w_{ij}^{(k+1)}+\frac{1}{\phi}t_{ij}^{(k)},\frac{\lambda_{2}}{\phi})$
        \EndFor
        \For{$i,j=1,\ldots,n,\:(i\neq j)$}
        \State $\bm{s}_{i,j}^{(k+1)} = \bm{s}_{i,j}^{(k)}+\rho(\bm{w}_{i}^{(k+1)}-\bm{a}_{i,j}^{(k+1)})$
        \EndFor
        \For{$i=1\ldots,n$}
        \State $\bm{t}_{i}^{(k+1)}=\bm{t}_{i}^{(k)}+\phi(\bm{w}_{i}^{(k+1)}-\bm{b}_{i}^{(k+1)})$
        \State $u_{i}^{(k+1)} = u_{i}^{(k)}+\psi\bm{1}_{p}^{T}\bm{w}_{i}^{(k+1)}$
        \EndFor
      \EndWhile
      \Ensure $\bm{b}_{i},\quad i=1,\ldots,n.$
    \end{algorithmic}
  \end{algorithm}
  
\begin{algorithm}[H]
    \caption{Estimation algorithm for constrained Weber problem (\ref{c-weber}) via ADMM}
    \label{Weber_ADMM}
    \begin{algorithmic}
      \Require Initialize\:$\bm{w}_{i^{\ast}}^{(0)},\bm{m}^{(0)},\bm{u}^{(0)},v^{(0)}$.
      \While{convergence}
        \For{$i=1\ldots,n$}
        \State $\bm{m}_{i}^{(k+1)} = \mathrm{min}\left(\frac{r_{i^{\ast},i}}{\mu},\|\bm{w}_{i^{\ast}}^{(k)}-\frac{1}{\mu}\bm{u}_{i}^{(k)} - \widehat{\bm{w}}_{i} \|_{2}\right)\frac{\bm{w}_{i^{\ast}}^{(k)}-\frac{1}{\mu}\bm{u}_{i}^{(k)} - \widehat{\bm{w}}_{i}}{\|\bm{w}_{i^{\ast}}^{(k)}-\frac{1}{\mu}\bm{u}_{i}^{(k)} - \widehat{\bm{w}}_{i} \|_{2}}$
        \EndFor
        \State $
        \bm{w}_{i^{\ast}}^{(k+1)} = ( \mu n I_{p}+\eta\bm{1}_{p}\bm{1}_{p}^{T} )^{-1}\left\{ \mu\sum_{i=1}^{n}(\bm{m}_{i}^{(k+1)}+\frac{1}{\mu}\bm{u}_{i}^{(k)})-v^{(k)}\bm{1}_{p}  \right\}$
        \For{$i=1\ldots,n$}
        \State $\bm{u}_{i}^{(k+1)} =\bm{u}_{i}^{(k)}+ \mu(\bm{m}_{i}^{(k+1)}-\bm{w}_{i^{\ast}}^{(k+1)})$
        \EndFor
        \State $v^{(k+1)} = v^{(k)} + \eta
\bm{1}_{p}^{T}\bm{w_{i^{\ast}}}^{(k+1)}$
      \EndWhile
      \Ensure $\bm{w}_{i^{\ast}}$
    \end{algorithmic}
  \end{algorithm}

\section{Simulation studies}\label{sec5}

In this section, we report simulation studies conducted with our proposed method using artificial data.
    
   In our simulations, we generated artificial data from the true model

        \begin{equation}
            y_{i} = 
            \begin{cases}
              \bm{z}_{i}^{T}\bm{w}_{(1)}^{\ast} + \epsilon_{i}, & (i=1,\ldots,40), \\
              \bm{z}_{i}^{T}\bm{w}_{(2)}^{\ast} + \epsilon_{i}, & (i=41,\ldots,80),\\
              \bm{z}_{i}^{T}\bm{w}_{(3)}^{\ast} + \epsilon_{i}, & (i=81,\ldots,120), 
            \end{cases}
        \end{equation}
        where $\bm{z}_{i}=(\log x_{i1},\ldots,\log x_{ip})^{T}$, $\bm{x}_{i}=(x_{i1},\ldots,x_{ip})^{T}$ is $p$-dimensional compositional data, $\bm{w}_{(1)}^{\ast}$,$\bm{w}_{(2)}^{\ast}$,$\bm{w }_{(3)}^{\ast}\in\mathbb{R}^{p}$ are the true regression coefficient vectors, and $\epsilon_{i}$ is an error term distributed as $N(0,0.1^{2})$ independently. We generated compositional data $\{\bm{x}_{i}; i=1,\ldots,120 \}$ as follows.
        First, we generated the data $\{\bm{c}_{i}, i=1,\ldots,120\}$ from the $p$-dimensional multivariate normal distribution $N_{p}(\bm{\omega},\Sigma)$ independently, where $(\bm{\omega})_{j}=\omega_{j}$, $(\Sigma)_{ij} = 0.2^{\abs{i-j}}$, and
        \begin{equation}
          \omega_{j} =
          \begin{cases}
            \log(0.5p), & (j=1,\ldots,5),\\
            0, & (j=6,\ldots,p).
          \end{cases} 
        \end{equation}
      Then, the data $\{\bm{c}_{i}, i=1,\ldots,120\}$ were converted to the compositional data $\{\bm{x}_{i}; i=1,\ldots,120 \}$ by transformation
        \begin{equation}
        \label{trans}
           x_{ij}=\frac{\exp(c_{ij})}{\sum_{k=1}^{p}\exp(c_{ik})},\quad i=1,\ldots,120,\: j=1,\ldots,p.
        \end{equation}
      The true regression coefficient vectors were set as
        \begin{equation}
            \begin{split}
                \bm{w}_{(1)}^{\ast}&=(1,-0.8,0.6,0,0,-1.5,-0.5,1.2,\bm{0}_{p-8}^{T})^{T}, \\ 
                \bm{w}_{(2)}^{\ast}&=(0,-0.5,1,1.2,0.1,-1,0,-0.8,\bm{0}_{p-8}^{T})^{T}, \\
                \bm{w}_{(3)}^{\ast}&=(0,0,0,0.8,1,0,-0.8,-1,\bm{0}_{p-8}^{T})^{T}.
            \end{split}
        \end{equation}
        We also assumed that the graph $R\in\{0,1\}^{120\times120}$ was observed. In the graph, the true value of each element was obtained with probability $P_{R}$. We calculated MSE as $\sum_{i^{\ast}=1}^{n^{\ast}}(y_{i^\ast}-\bm{z}_{i^{\ast}}^{T}\widehat{\bm{w}}_{i^{\ast}})^{2}/n^{\ast}$, dividing the 120 samples into 100 training data and 20 validation data. Here, $n^{\ast}$ indicates the number of samples for validation data (i.e., 20). The regression coefficient vectors for the validation data were estimated based on constrained Weber problem (\ref{c-weber}). To evaluate the effectiveness of our proposed method, we compared it with both Model (\ref{syml1}) and the model obtained by removing the zero-sum constraint from minimization problem (\ref{prop}). We refer to the latter two comparison methods as CL and SNL, respectively. The regularization parameters were determined by five-fold CV for both the proposed method and the comparison methods. The values of tuning parameters $\rho,\phi,\psi,\mu,\eta$ for ADMM were all set to one. We considered several settings: $p=\{30, 100, 200\}$, $P_{R}=\{0.99, 0.95, 0.90, 0.80, 0.70\}$. We generated 100 datasets and computed the mean and standard deviation of MSE from the 100 repetitions.

    Table \ref{simresult} shows the results for the mean and standard deviation of MSE. The proposed method and SNL show better prediction accuracy than CL in all settings. The reason for this may be that CL assumes only a single regression coefficient vector and thus fails to capture the true structure, which consists of three clusters. A comparison between the proposed method and SNL shows that the proposed method has higher prediction accuracy than SNL when $P_{R}=0.99, 0.95,$ and $0.90$, whereas SNL shows better results in most cases when $P_{R}=0.80, 0.70$. This means that the proposed method is superior to SNL when the structure of the graph $R$ is similar to the true structure. On the whole, prediction accuracy deteriorates as $P_{R}$ decreases for both the proposed method and SNL, but this deterioration is more extreme for the proposed method. For both the proposed method and SNL, which assume multiple regression coefficient vectors, standard deviation is somewhat large.
        
    \begin{table}[ht]
        \centering
        \caption{Mean and deviation of MSE for simulations.}
        \label{simresult}
        \begin{tabular}{cccc}
            \hline
            Method   & $p=30$ & $p=100$ & $p=200$ \\ \hline
            CL  & $5.20(1.44)$ & $6.99(1.81)$ & $8.75(2.69)$ \\ \hline
            \multicolumn{4}{c}{$P_{R}=0.99$}
 \\ 
            Proposed & $\bm{0.51}(0.58)$ & $\bm{2.13}(2.35)$ & $\bm{2.70}(1.92)$ \\
            SNL & $2.54(0.97)$ & $3.73(1.31)$ & $6.55(3.94)$ \\ \hline
            \multicolumn{4}{c}{$P_{R}=0.95$} \\ 
            Proposed & $\bm{2.74}(1.19)$ & $\bm{2.96}(1.33)$ & $\bm{3.68}(2.71)$ \\
            SNL & $3.19(0.98)$ & $3.87(1.35)$ & $5.26(4.15)$ \\ \hline
            \multicolumn{4}{c}{$P_R=0.90$} \\ 
            Proposed & $\bm{3.29}(1.38)$ & $\bm{3.80}(1.33)$ & $\bm{4.49}(1.81)$ \\
            SNL & $3.40(1.23)$ & $4.25(1.49)$ & $4.75(1.49)$ \\ \hline
            \multicolumn{4}{c}{$P_R=0.80$} \\ 
            Proposed & $4.20(1.58)$ & $5.53(2.30)$ & $7.49(3.98)$ \\ 
            SNL & $\bm{3.87}(1.41)$ & $\bm{4.86}(1.52)$ & $\bm{5.70}(2.00)$ \\ \hline
            \multicolumn{4}{c}{$P_R=0.70$} \\ 
            Proposed & $\bm{4.13}(1.56)$ & $6.57(2.55)$ & $7.66(2.28)$ \\ 
            SNL & $4.56(1.55)$ & $\bm{5.63}(1.72)$ & $\bm{6.69}(2.25)$ \\ \hline
        \end{tabular}
    \end{table}

\section{Application to gut microbiome data}\label{sec6}

    The gut microbiome affects human health/disease, for example, in terms of obesity. Gut microbiomes may be exposed from inter-individual heterogeneity from environmental factors such as diet, as well as from hormonal factors and age (\cite{Haro2016-xj}; \cite{Saraswati2015-cg}). In this section, we applied our proposed method to the real dataset reported by \cite{Wu2011-os}. 
This dataset is from a cross-sectional study of 98 healthy volunteers conducted at the University of 
Pennsylvania to investigate the connections between long-term dietary patterns and gut microbiome composition. 
Stool samples were collected from the subjects, and DNA samples were
analyzed by 454/Roche pyrosequencing of 16S rRNA gene segments of the V1--V2 region. In the results, the counts for more than 17,000 species-level OTUs were obtained. 
Demographic data such as body mass index (BMI), sex, and age were also obtained.
   
   We used centered BMI as the response and the compositional data of gut microbiome as the explanatory variable. In order to reduce the number of the OTUs, we used the single-linkage clustering based on an available phylogenetic tree to aggregate the OTUs, which is provided as the function \textbf{tip\_glom} in the R package ``phyloseq" (see \cite{McMurdie2013-do}).  
In this process, some closely related OTUs defined on the leaf nodes of the phylogenetic tree are aggregated into one OTU when the cophenetic distances between the OTUs are smaller than a threshold. We set the threshold at $0.5$. As a result, 166 OTUs were obtained by the aggregation. Since some OTUs had zero counts making it impossible to take the logarithm, these counts were replaced by the value one before converting them to compositional data. 
   
   We computed the graph $R\in\mathbb{R}^{98\times 98}$ by
     \begin{equation}
            R = \frac{S^{T}+S}{2}, \quad (S)_{ij}=
            \begin{cases}
                 1 & \mbox{$j$-th sample is a 5-NN of $i$-th sample with}\;D_{ij}, \\
                 0 & \mbox{Otherwise},
            \end{cases}
        \end{equation}
    where $D_{ij}$ is the distance between the $i$-th and $j$-th samples. Distance $D_{ij}$ was calculated in the following two ways. (i) Gower distance \citep{Gower1971-zp} was calculated using the sex and age data of the subjects. (ii) Log-ratio distance (e.g., see \cite{Greenacre2018-dn}) was calculated using the explanatory variable, as follows:
    \begin{equation}
        D_{ij}= \sqrt{\sum_{l=1}^{p}(z_{il}^{c}-z_{jl}^{c})^{2}},
    \end{equation}
  where $z^{c}_{ij}=\log x_{ij}-\frac{1}{p}\sum_{j=1}^{p}\log x_{ij}$. Using these two ways of calculating distance, we obtained two different $R$ and estimation results. We refer to these two ways as Proposed (i) and Proposed (ii), respectively.

   To evaluate the prediction accuracy of our proposed method, we calculated MSE by randomly splitting the dataset into 90 samples as the training data and 8 samples as the validation data. We again used the method of Lin et al. (2014) referred to as CL as a comparison method. The regularization parameters were determined by five-fold CV for both our proposed method and CL. The mean and standard deviation of MSE were calculated from 100 repetitions.
    
    Table \ref{real_PE} shows the mean and standard deviation of MSE in the real data analysis. We observe that Proposed (i) has the smallest mean and standard deviation of MSE. This indicates that our proposed method captures the inherent structure of the compositional data by providing an appropriate graph $R$. However, standard deviation is large even for Proposed (i), which indicates that prediction accuracy may strongly depend on what method is used to split the dataset.
    
    \begin{table}[ht]
    \centering
    \caption{Mean and standard deviation of MSE for real data analysis (100 repetitions)}       
    \label{real_PE}
        \begin{tabular}{cccc}
                    \hline
               Value  & Proposed (i) & Proposed (ii) & CL \\ \hline
                MSE (SD) & $\bm{23.01} (16.62)$ & $31.59 (22.44)$ & $30.96 (23.36)$ \\ \hline
        \end{tabular}
    \end{table}
    
\begin{figure}[ht]
    \begin{minipage}[]{0.5\linewidth}
        \centering
        \includegraphics[width=4.8cm,height=4.8cm]{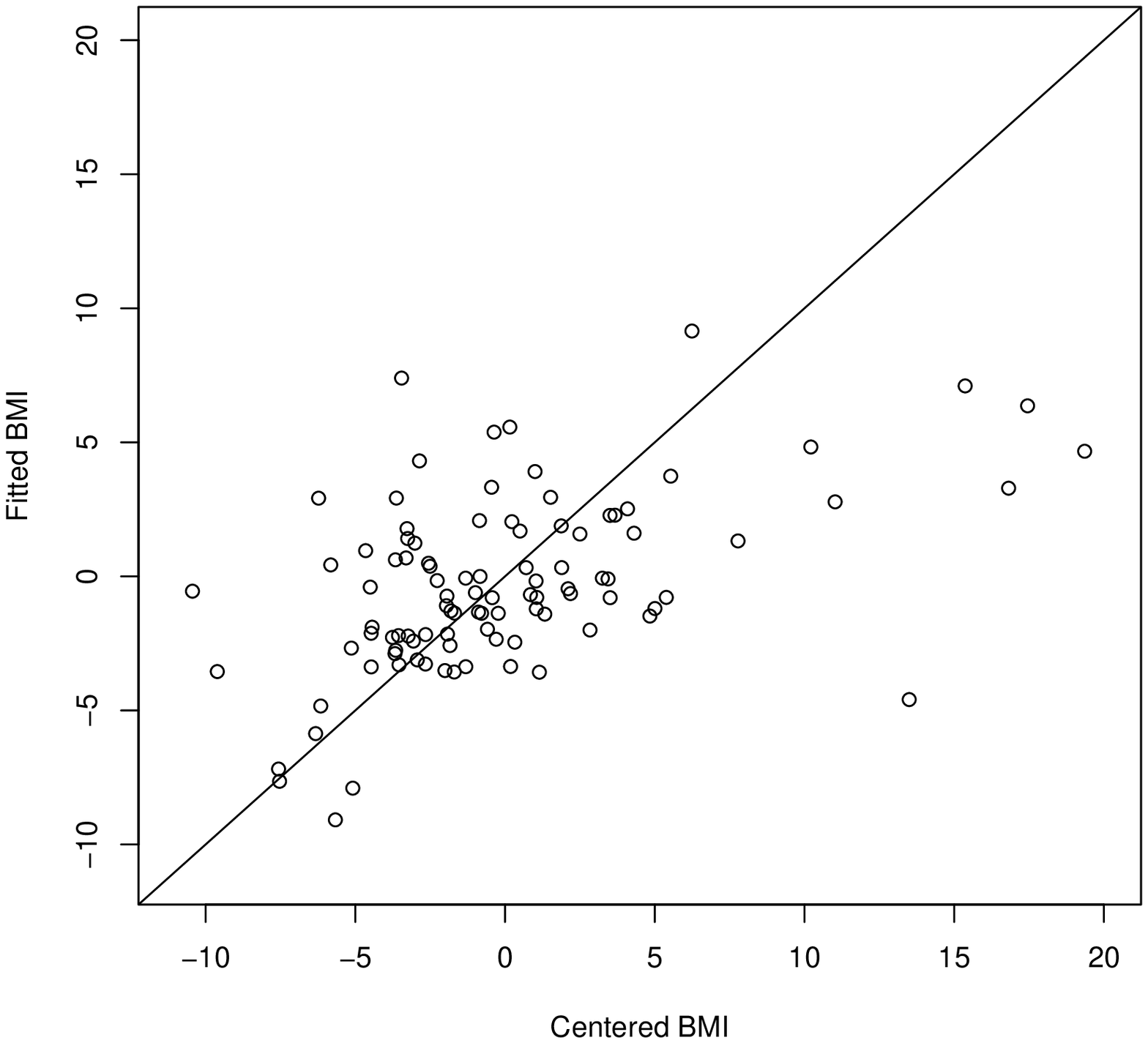} 
        \vspace{-3mm}
        \subcaption{Proposed (i)}
        \label{fig:Proposed}
        \end{minipage}
        \begin{minipage}[]{0.5\linewidth}
        \centering
        \includegraphics[width=4.8cm,height=4.8cm]{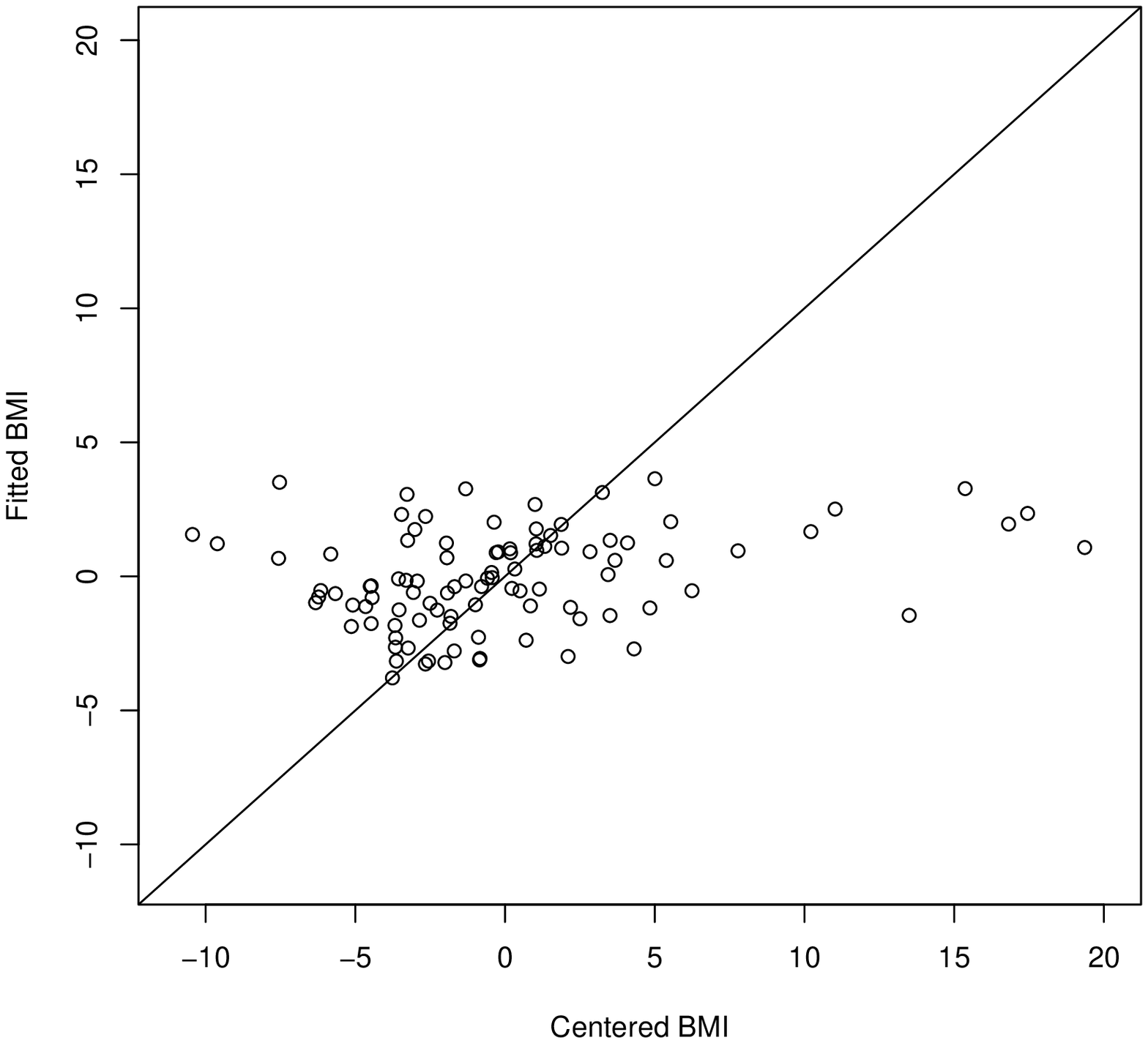} 
        \vspace{-3mm}
        \subcaption{CL}
        \label{fig:CL}
        \end{minipage}
        \caption{Observed and predicted BMI using LOOCV.} 
        \label{fig:ObservedPredictedLOOCV}
    \end{figure}

    Table \ref{real_R2} shows the coefficient of determination $\mathrm{R}^2$ using leave-one-out cross-validation (LOOCV) for Proposed (i) and CL. The fittings of the observed and predicted BMI are plotted in Figures \ref{fig:ObservedPredictedLOOCV} \subref{fig:Proposed} and \subref{fig:CL} for Proposed (i) and CL, respectively. The horizontal axis represents the centered observed BMI values and the vertical axis represents the corresponding predicted BMI. As shown, CL does not predict data with centered observed values between $-10$ and $-5$ as being in that interval, whereas Proposed (i) predicts these data correctly to some extent.
     \begin{table}[ht] 
        \centering
        \caption{Coefficients of determination using LOOCV}
        \label{real_R2}
        \begin{tabular}{ccc}
                    \hline
               Value  & Proposed (i) & CL \\ \hline
                LOOCV $\mathrm{R}^{2}$ & $\bm{0.245}$ & $0.083$  \\ \hline
        \end{tabular}
    \end{table}

\begin{figure}[ht]
        \centering
        \includegraphics[width=10cm]{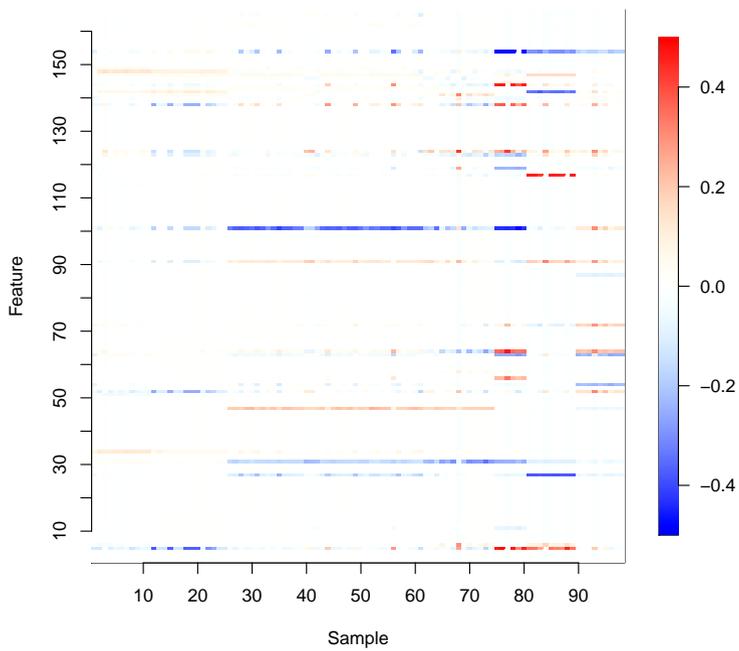}
        \caption{Estimated regression coefficients for all samples.}
        \label{Wsort}
    \end{figure}

    Figure \ref{Wsort} shows the regression coefficients estimated by Proposed (i) for all samples, where the regularization parameters are determined by LOOCV. For the results in Figure \ref{Wsort}, we used hierarchical clustering to group together similar regression coefficient vectors, setting the number of clusters as five.
    
\begin{figure}[ht]
        \centering
        \includegraphics[width=12cm]{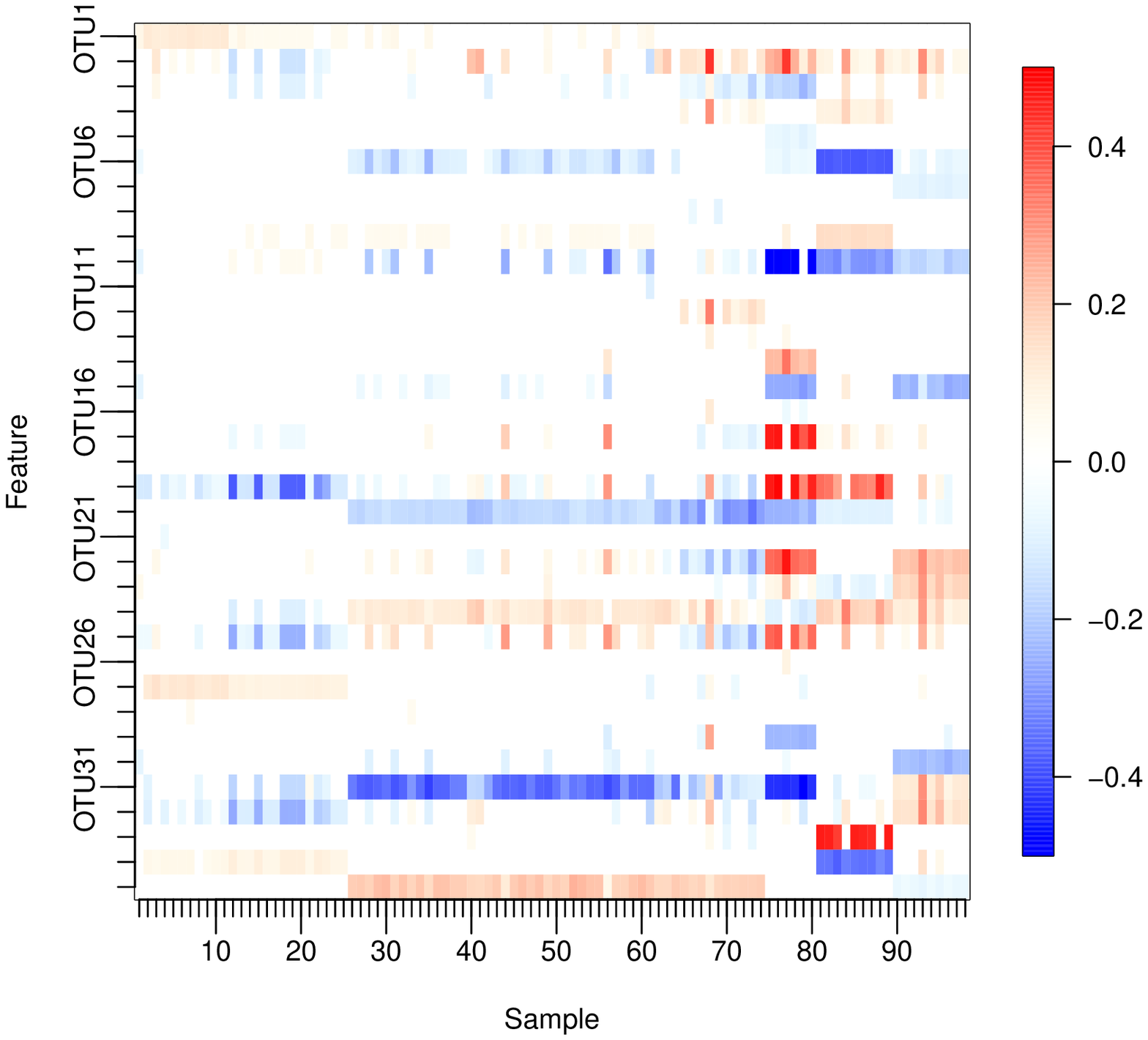}
        \caption{Only the estimated regression coefficients with $\abs{\widehat{w}_{ij}}\geq0.05$ for at least one sample.}
        \label{Wsortleft}
    \end{figure}

With our proposed method, many of estimated regression coefficients were not exactly zero but close to zero. Thus, we will treat estimated regression coefficients  $\abs{\widehat{w}_{ij}}<0.05$ as being exactly zero to simplify the interpretation.
 Figure \ref{Wsortleft} shows only those coefficients that satisfy $\abs{\widehat{w}_{ij}}\geq0.05$ in at least one sample,
where the corresponding variables are listed in Table \ref{Wsortname2}.
 
  It is reported that the human gut microbiome can be classified into three clusters called enterotypes, which are characterized by three predominant genera, \textit{Bacteroides}, \textit{Prevotella}, and \textit{Ruminococcus} \citep{Arumugam2011-si}. In the dataset, OTUs of genus-levels \textit{Prevotella} and \textit{Ruminococcus} were aggregated into the OTUs of family-levels \textit{Prevotellaceae} and \textit{Ruminococcaceae} by the single-linkage clustering. In Figure \ref{Wsortleft}, \textit{Bacteroides} correspond to OTU5, 6, 7, 8, 9, and 10, \textit{Prevotellaceae} to OTU12, and \textit{Ruminococcaceae} to OTU30 and 31.
  For these OTUs, the differences are clear between OTU6, 9, 10 and OTU30, 31 among samples 81--90, in which only \textit{Bacteroides} are correlated to the response. On the other hand, the differences among samples 65--74 are also indicated, in which only \textit{Bacteroides} do not affect BMI. These results suggest that \textit{Bacteroides}, \textit{Prevotellaceae}, and \textit{Ruminococcaceae} may have different effects on BMI that are associated with enterotypes.
  In addition, it is reported that women with a higher abundance of \textit{Prevotellaceae} are more obese \citep{Cuevas-Sierra2020-wy}. The regression coefficients of non-zero \textit{Prevotellaceae} are all positive, and the corresponding 8 samples are all females. On the other hand, in OTU29 indicating \textit{Roseburia}, 9 samples out of 10 are negatively associated with BMI. \textit{Roseburia} is also reported to be negatively correlated with indicators of bodyweight \citep{Zeng2019-nx}. 
   
\begin{landscape}
 
\begin{table}[ht]
    \centering
    \caption{Variables with estimated regression coefficients $\abs{\widehat{w}_{ij}}\geq0.05$ for at least one sample.}
    \label{Wsortname2}
    \small
    \scalebox{0.8}{
    \begin{tabular}{cccccccc}
    \hline
      Variable   & Kingdom & Phylum & Class & Order & Family & Genus & Species  \\ \hline
        OTU1 & \textit{Bacteria} & & & & & & \\ 
        OTU2 & \textit{Bacteria} &  &  &  &  &  &  \\ 
        OTU3 & \textit{Bacteria} & \textit{Bacteroidetes}  &  &  &  &  &  \\ 
        OTU4 & \textit{Bacteria} & \textit{Bacteroidetes} & \textit{Bacteroidetes} & \textit{Bacteroidales}  &  &  &  \\ 
        OTU5 & \textit{Bacteria} & \textit{Bacteroidetes} & \textit{Bacteroidetes} & \textit{Bacteroidales} & \textit{Bacteroidaceae} & \textit{Bacteroides} &  \\ 
        OTU6 & \textit{Bacteria} & \textit{Bacteroidetes} & \textit{Bacteroidetes} & \textit{Bacteroidales} & \textit{Bacteroidaceae} & Bacteroides &  \\ 
        OTU7 & \textit{Bacteria} & \textit{Bacteroidetes} & \textit{Bacteroidetes} & \textit{Bacteroidales} & \textit{Bacteroidaceae} & \textit{Bacteroides} &  \\ 
        OTU8 & \textit{Bacteria} & \textit{Bacteroidetes} & \textit{Bacteroidetes} & \textit{Bacteroidales} & \textit{Bacteroidaceae} & \textit{Bacteroides} &  \\ 
        OTU9 & \textit{Bacteria} & \textit{Bacteroidetes} & \textit{Bacteroidetes} & \textit{Bacteroidales} & \textit{Bacteroidaceae} & \textit{Bacteroides} &  \\ 
        OTU10 & \textit{Bacteria} & \textit{Bacteroidetes} & \textit{Bacteroidetes} & \textit{Bacteroidales} & \textit{Bacteroidaceae} & \textit{Bacteroides} &  \\ 
        OTU11 & \textit{Bacteria} & \textit{Bacteroidetes} & \textit{Bacteroidetes} & \textit{Bacteroidales} & \textit{Porphyromonadaceae} & \textit{Parabacteroides}  &  \\ 
        OTU12 & \textit{Bacteria} & \textit{Bacteroidetes} & \textit{Bacteroidetes} & \textit{Bacteroidales} & \textit{Prevotellaceae} &  &  \\ 
        OTU13 & \textit{Bacteria} & \textit{Firmicutes} & \textit{Clostridia} &  &  &  &  \\ 
        OTU14 & \textit{Bacteria} & \textit{Firmicutes} & \textit{Clostridia} & \textit{Clostridiales} &  &  &  \\ 
        OTU15 & \textit{Bacteria} & \textit{Firmicutes} & \textit{Clostridia} & \textit{Clostridiales} &  &  &  \\ 
        OTU16 & \textit{Bacteria} & \textit{Firmicutes} & \textit{Clostridia} & \textit{Clostridiales} &  &  &  \\ 
        OTU17 & \textit{Bacteria} & \textit{Firmicutes} & \textit{Clostridia} & \textit{Clostridiales} &  &  &  \\ 
        OTU18 & \textit{Bacteria} & \textit{Firmicutes} & \textit{Clostridia} & \textit{Clostridiales} &  &  &  \\ 
        OTU19 & \textit{Bacteria} & \textit{Firmicutes} & \textit{Clostridia} & \textit{Clostridiales} & \textit{Lachnospiraceae} &  &  \\ 
        OTU20 & \textit{Bacteria} & \textit{Firmicutes} & \textit{Clostridia} & \textit{Clostridiales} & \textit{Lachnospiraceae} &  &  \\ 
        OTU21 & \textit{Bacteria} & \textit{Firmicutes} & \textit{Clostridia} & \textit{Clostridiales} & \textit{Lachnospiraceae} &  &  \\ 
        OTU22 & \textit{Bacteria} & \textit{Firmicutes} & \textit{Clostridia} & \textit{Clostridiales} & \textit{Lachnospiraceae} &  &  \\ 
        OTU23 & \textit{Bacteria} & \textit{Firmicutes} & \textit{Clostridia} & \textit{Clostridiales} & \textit{Lachnospiraceae} &  &  \\ 
        OTU24 & \textit{Bacteria} & \textit{Firmicutes} & \textit{Clostridia} & \textit{Clostridiales} & \textit{Lachnospiraceae} &  &  \\ 
        OTU25 & \textit{Bacteria} & \textit{Firmicutes} & \textit{Clostridia} & \textit{Clostridiales} & \textit{Lachnospiraceae} & &  \\ 
        OTU26 & \textit{Bacteria} & \textit{Firmicutes} & \textit{Clostridia} & \textit{Clostridiales} & \textit{Lachnospiraceae} &  &  \\ 
        OTU27 & \textit{Bacteria} & \textit{Firmicutes} & \textit{Clostridia} & \textit{Clostridiales} & \textit{Lachnospiraceae} &  &  \\ 
        OTU28 & \textit{Bacteria} & \textit{Firmicutes} & \textit{Clostridia} & \textit{Clostridiales} & \textit{Lachnospiraceae} &  &  \\ 
        OTU29 & \textit{Bacteria} & \textit{Firmicutes} & \textit{Clostridia} & \textit{Clostridiales} & \textit{Lachnospiraceae} & \textit{Roseburia} &  \\ 
        OTU30 & \textit{Bacteria} & \textit{Firmicutes} & \textit{Clostridia} & \textit{Clostridiales} & \textit{Ruminococcaceae} & &   \\ 
        OTU31 & \textit{Bacteria} & \textit{Firmicutes} & \textit{Clostridia} & \textit{Clostridiales} & \textit{Ruminococcaceae} &  &  \\ 
        OTU32 & \textit{Bacteria} & \textit{Firmicutes} & \textit{Erysipelotrichia} & \textit{Erysipelotrichales} & \textit{Erysipelotrichaceae} & \textit{Catenibacterium} &  \\ 
        OTU33 & \textit{Bacteria} & \textit{Firmicutes} & \textit{Erysipelotrichia} & \textit{Erysipelotrichales} & \textit{Erysipelotrichaceae} & \textit{Erysipelotrichaceae.Incertae.Sedis} &  \\ 
        OTU34 & \textit{Bacteria} & \textit{Proteobacteria} &  &  &  &  &  \\ 
        OTU35 & \textit{Bacteria} & \textit{Proteobacteria} & \textit{Gammaproteobacteria} & \textit{Enterobacteriales} & \textit{Enterobacteriaceae} &  &  \\ \hline
    \end{tabular}
    }
\end{table}
\end{landscape}
\section{Conclusion}

We proposed a multi-task learning method for compositional data based on a network lasso and log-contrast model. 
By imposing a zero-sum constraint on the model corresponding to each sample, we could extract the information of latent clusters in the regression coefficient vectors for compositional data. In the results of simulations, the proposed method worked well when clusters existed for the compositional data and an appropriate graph $R$ was obtained. 
In a human gut microbiome analysis, our proposed method provided high prediction accuracy compared with the existing method by considering the heterogeneity from age and sex. In addition, cluster-specific OTUs such as ones related to enterotypes were detected in terms of effects on BMI.

Although our proposed method shrinks some regression coefficients that do not affect response to zero, many coefficients close to zero remain.
Furthermore, in both the simulations and human gut microbiome analysis, the prediction accuracy of the proposed method deteriorated significantly when the obtained $R$ did not capture the true structure. Moreover, the standard deviations of MSE were high in almost all cases.
We leave these as topics of future research. 

\backmatter
\bmhead{Acknowledgments}
S. K. was supported by JSPS KAKENHI Grant Numbers JP19K11854 and JP20H02227. Supercomputing resources were provided by the Human Genome Center (the Univ. of Tokyo).
\noindent

\bigskip

\begin{appendices}
\section{Derivations of update formulas in ADMM}
\subsection{\texorpdfstring{Update of $\bm{w}$}{wij}}
In the update of $\bm{w}$, we minimize the terms of the augmented Lagrangian (\ref{lagrange}) depending on $\bm{w_{i}}$ as follows:
\begin{equation}
    \begin{split}
        \bm{w}_{i}^{(k+1)} = \argmin_{\bm{w}_{i}\in\mathbb{R}^{p}} \left\{ (y_{i} - \bm{w}_{i}^{T}\bm{z}_{i})^{2}+\frac{\rho}{2}\sum_{j\neq i}^{n}\|\bm{w}_{i}-\bm{a}_{i,j}^{(k)}+\bm{s}_{i,j}^{(k)} \|_{2}^{2} \right. \\
\left.      +\bm{t}_{i}^{(k)T}(\bm{w}_{i}-\bm{b}_{i}^{(k)})+\frac{\phi}{2}\|\bm{w}_{i}-\bm{b}_{i}^{(k)} \|_{2}^{2}+u_{i}^{(k)}\bm{1}_{p}^{T}\bm{w}_{i} + \frac{\psi}{2}\|\bm{1}_{p}^{T}\bm{w}_{i}\|_{2}^{2} \right\}.
    \end{split}
\end{equation}
 From $\frac{\partial L}{\partial \bm{w}_{i}} = \bm{0}$, we obtain the update
\begin{equation}
    \begin{split}
        \bm{w}_{i}^{(k+1)} &= \left\{2\bm{z}_{i}\bm{z}_{i}^{T}+(\rho(n-1)+\phi)I_{p}+\psi\bm{1}_{p}\bm{1}_{p}^{T}\right\}^{-1} \\
        &\left\{2y_{i}\bm{z}_{i}+\rho\sum_{j\neq i}^{n}(\bm{a}_{i,j}^{(k)}-\bm{s}_{i,j}^{(k)})-\bm{t}_{i}^{(k)}+\phi\bm{b}_{i}^{(k)}-u_{i}^{(k)}\bm{1}_{p}\right\}
    \end{split}
\end{equation}
    
    \subsection{\texorpdfstring{Update of $\bm{a}$}{a}}
The update of $\bm{a}$ is obtained by joint minimization of $\bm{a}_{i,j}$ and $\bm{a}_{j,i}$ as follows:
        \begin{equation}
            \begin{split}
             \bm{a}_{i,j}^{(k+1)},\bm{a}_{j,i}^{(k+1)} &= \argmin_{\bm{a}_{i,j},\bm{a}_{j,i}\in \mathrm{R}^{p}} \left\{\lambda_{1} r_{i,j}\|a_{i,j}-a_{j,i} \|_{2} \right. \\ &+ \left. \frac{\rho}{2}(\|\bm{w}_{i}^{(k+1)}-\bm{a}_{i,j}+\bm{s}_{i,j}^{(k)} \|_{2}^{2}+\|\bm{w}_{j}^{(k+1)}-\bm{a}_{j,i}+\bm{s}_{j,i}^{(k)}\|_{2}^{2}) \right\}
            \end{split}
        \end{equation}
 In Hallac et al. (2011), the analytical solution was given as 
        \begin{equation}
            \begin{split}
                \bm{a}_{ij}^{(k+1)}=\theta(\bm{w}_{i}^{(k+1)}+\bm{s}_{i,j}^{(k)})+(1-\theta)(\bm{w}_{j}^{(k+1)}+\bm{s}_{j,i}^{(k)}), \\
                \bm{a}_{ji}^{(k+1)}=(1-\theta)(\bm{w}_{i}^{(k+1)}+\bm{s}_{i,j}^{(k)})+\theta(\bm{w}_{j}^{(k+1)}+\bm{s}_{j,i}^{(k)}),
            \end{split}
        \end{equation}
            where
        \begin{equation}
            \theta ={\rm max}\left(1-\frac{\lambda_{1} r_{i,j}}{\rho\|(\bm{w}_{i}^{(k+1)}+\bm{s}_{i,j}^{(k)})-(\bm{w}_{j}^{(k+1)}+\bm{s}_{j,i}^{(k)}) \|_{2}},0.5 \right).
        \end{equation}
        
    \subsection{\texorpdfstring{Update of $\bm{b}$}{b}}
    The update of $\bm{b}$ is obtained by the minimization problem
    \begin{equation}
        \bm{b}_{i}^{(k+1)} = \argmin_{\bm{b}_{i}\in \mathrm{R}^{p}}  \sum_{i=1}^{n} \left\{ \lambda_{2}\|\bm{b}_{i} \|_{1}+\bm{t}_{i}^{T}(\bm{w}_{i}-\bm{b}_{i})+\frac{\phi}{2}\|\bm{w}_{i}-\bm{b}_{i} \|_{2}^{2} \right\}.
    \end{equation}
   Because the minimization problem with respect to $\bm{b}_{i}$ contains a non-differentiable point in the $\ell_{1}$ norm of $\bm{b}_{i}$, we consider the subderivative of $\abs{b_{ij}}$. Then we obtain the update
     \begin{equation}
      b_{ij}^{(k+1)} =  \mathrm{S}\left( w_{ij} + \frac{t_{i,j}}{\phi}, \frac{\lambda_{2}}{\phi} \right),\quad j=1,\ldots,p,
    \end{equation}
    where $\mathrm{S}(\cdot,\cdot)$ is the soft-thresholding operator given by $\mathrm{S}(x,\lambda) := {\rm sign}(x)(\abs{x}-\lambda)_{+}$.
    \subsection{\texorpdfstring{Update of $Q$}{Q}}
     The updates for the Lagrange multipliers denoted as $Q$ are obtained by gradient descent as follows:
    \begin{equation}
        \begin{split}
       \bm{s}_{i,j}^{(k+1)} &= \bm{s}_{i,j}^{(k)}+\rho(\bm{w}_{i}^{(k+1)}-\bm{a}_{i,j}^{(k+1)}) \quad i,j=1,\ldots,n\;(i\neq j),\\
      \bm{t}_{i}^{(k+1)}&=\bm{t}_{i}^{(k)}+\phi(\bm{w}_{i}^{(k+1)}-\bm{b}_{i}^{(k+1)}) \quad i=1,\ldots,n,\\
    u_{i}^{(k+1)} &= u_{i}^{(k)}+\psi\bm{1}_{p}^{T}\bm{w}_{i}^{(k+1)} \quad i=1,\ldots,n.
        \end{split}
    \end{equation}
    
\section{Update algorithm for constrained Weber problem via ADMM}
   We consider the updates for the following constrained Weber problem via ADMM based on \cite{Chaudhury2015-wp}.
 \begin{equation}
        \label{c-weber2}
        \min_{\bm{w}_{i^{\ast}}\in \mathbb{R}^{p}}  \left\{ \sum_{i=1}^{n} r_{i^{\ast},i}  \| \bm{w}_{i^{\ast}} - \widehat{\bm{w}}_{i} \|_{2} \right\}\quad{\mathrm{s.t.}}\quad \sum_{j=1}^{p}w_{i^{\ast}j}=0.
    \end{equation}
    Minimization problem (\ref{c-weber2}) is equivalently represented as
    
    \begin{equation}
        \label{c-weber-ADMM}
        \begin{split}
            \min_{\bm{w}_{i^{\ast}},\bm{m}_{1},\ldots,\bm{m}_{n}\in \mathbb{R}^{p}}  \left\{ \sum_{i=1}^{n} r_{i^{\ast},i}  \| \bm{m}_{i} - \widehat{\bm{w}}_{i} \|_{2} \right\}, \\
            \mathrm{s.t.} \quad\bm{1}_{p}^{T}\bm{w}_{i^{\ast}} = 0, \:\bm{m}_{i} = \bm{w}_{i^{\ast}},\quad i=1,\ldots,n.
        \end{split}
    \end{equation}
    The augmented Lagrangian for (\ref{c-weber-ADMM}) is formulated as
    \begin{equation}
        \label{aug-weber}
        \begin{split}
           L_{\rho,\phi}(\bm{w}_{i^\ast},\bm{m}_{1},\ldots,\bm{m}_{n})&= \sum_{i=1}^{n} \left\{r_{i^{\ast},i}  \| \bm{m}_{i} - \widehat{\bm{w}}_{i} \|_{2}+\bm{u}_{i}^{T}(\bm{m}_{i}-\bm{w}_{i^{\ast}})+\frac{\mu}{2}\|\bm{m}_{i}-\bm{w}_{i^{\ast}}\|_{2}^{2} \right\} \\
           &+v\bm{1}_{p}^{T}\bm{w}_{i^{\ast}}+\frac{\eta}{2}\|\bm{1}_{p}^{T}\bm{w}_{i^{\ast}}\|_{2}^{2},
        \end{split}
    \end{equation}
    where $\bm{u}_{i},v$ are Lagrange multipliers and $\mu,\eta\:(>0)$ are tuning parameters.
    \subsection{\texorpdfstring{Update of $\bm{w}_{i^{\ast}}$}{w}}
    In the update of $\bm{w}_{i^{\ast}}$, we minimize the terms of augmented Lagrangian (\ref{aug-weber}) depending on $\bm{w_{i^{\ast}}}$ as follows:
    \begin{equation}
        \bm{w}_{i^{\ast}}^{(k+1)} = \argmin_{w_{i^{\ast}}\in\mathbb{R}^{p}} \left\{ \sum_{i=1}^{n} \left\{\frac{\mu}{2} \left\|\bm{w}_{i^{\ast}}-\bm{m}_{i}^{(k)}-\frac{1}{\mu}\bm{u}_{i}^{(k)} \right\|_{2}^{2} \right\} \\
           +v^{(k)}\bm{1}_{p}^{T}\bm{w}_{i^{\ast}}+\frac{\eta}{2}\|\bm{1}_{p}^{T}\bm{w}_{i^{\ast}}\|_{2}^{2} \right\}
    \end{equation}
    From $\frac{\partial L}{\partial \bm{w}_{i^{\ast}}} = \bm{0}$, we obtain the update

    \begin{equation}
        \bm{w}_{i^{\ast}}^{(k+1)} = ( \mu n \bm{1}_{p}+\eta\bm{1}_{p}\bm{1}_{p}^{T} )^{-1}\left\{ \mu\sum_{i=1}^{n} \left(\bm{m}_{i}^{(k)}+\frac{1}{\mu}\bm{u}_{i}^{(k)} \right)-v^{(k)}\bm{1}_{p}  \right\}.
    \end{equation}
  
    \subsection{\texorpdfstring{Update of $\bm{m}$}{m}}
    In the update of $\bm{m}_{i}$, we minimize the terms of augmented Lagrangian (\ref{aug-weber}) depending on $\bm{m}_{i}$ as follows:
    \begin{equation}
      \label{update_m}
      \begin{split}
          \bm{m}_{i}^{(k+1)}=\argmin_{\bm{m}_{i}\in\mathbb{R}^{p}}\left\{ r_{i^{\ast},i}\|\bm{m}_{i}-\bm{w}_{i^{\ast}}^{(k+1)} \|_{2}+\frac{\mu}{2} \left\|\bm{m}_{i}- \left(\bm{w}_{i^{\ast}}^{(k+1)}-\frac{1}{\mu}\bm{u}_{i}^{(k)} \right) \right\|_{2}^{2} \right\},\quad \\ i=1,\dots,n,
      \end{split}
    \end{equation}
    Minimization problem (\ref{update_m}) is equivalently expressed as
    \begin{equation}
        \label{update_m2}
        \begin{split}
            \bm{m}_{i}^{(k+1)}=\argmin_{\bm{m}_{i}\in\mathbb{R}^{p}}\left\{ \frac{r_{i^{\ast},i}}{\mu} \|\bm{m}_{i}-\bm{w}_{i^{\ast}}^{(k)} \|_{2}+\frac{1}{2} \left\|\bm{m}_{i}-\left(\bm{w}_{i^{\ast}}^{(k)}-\frac{1}{\mu}\bm{u}_{i}^{(k)}\right) \right\|_{2}^{2} \right\}, \\ i=1,\dots,n.
        \end{split}
    \end{equation}
   In \cite{Chaudhury2015-wp}, because the right-hand side of (\ref{update_m2}) is the proximal map of the function  $f(\bm{m}_{i})=\frac{r_{i^{\ast},i}}{\rho}\|\bm{m}_{i}-\bm{w}_{i^{\ast}} \|_{2}$ by using Moreau's decomposition (e.g., \cite{Parikh2014-hd}), the updates of $\bm{m}$ are obtained by
    \begin{equation}
        \bm{m}_{i}^{(k+1)} = \min\left(\frac{r_{i^{\ast},i}}{\mu},\left\|\bm{w}_{i^{\ast}}^{(k)}-\frac{1}{\mu}\bm{u}_{i}^{(k)} - \widehat{\bm{w}}_{i} \right\|_{2}\right)\frac{\bm{w}_{i^{\ast}}^{(k)}-\frac{1}{\mu}\bm{u}_{i}^{(k)} - \widehat{\bm{w}}_{i}}{\|\bm{w}_{i^{\ast}}^{(k)}-\frac{1}{\mu}\bm{u}_{i}^{(k)} - \widehat{\bm{w}}_{i} \|_{2}}.
    \end{equation}
    
    \subsection{\texorpdfstring{Update of $\bm {u}$ and $v$}{u,v}}
    The updates for Lagrange multipliers $\bm{u}_{i}$ and $v$ are obtained by gradient descent as follows:
    \begin{equation}
        \begin{split}
            \bm{u}_{i}^{(k+1)} &=\bm{u}_{i}^{(k)}+ \mu(\bm{m}_{i}^{(k+1)}-\bm{w}_{i^{\ast}}^{(k+1)}), \quad i=1,\ldots,n, \\
            v^{(k+1)} &= v^{(k)} + \eta\bm{1}_{p}^{T}\bm{w_{i^{\ast}}}^{(k+1)}.
        \end{split}
    \end{equation}
    


\end{appendices}


\bibliography{Paperpile.bib}

\begin{thebibliography}{28}
\providecommand{\natexlab}[1]{#1}
\providecommand{\url}[1]{{#1}}
\providecommand{\urlprefix}{URL }
\providecommand{\doi}[1]{\url{https://doi.org/#1}}
\providecommand{\eprint}[2][]{\url{#2}}
 \bibcommenthead

\bibitem[{Aitchison(1982)}]{Aitchison1982-nl}
Aitchison J (1982) The statistical analysis of compositional data. Journal of
  the Royal Statistical Society Series B 44(2):139--160

\bibitem[{Aitchison and Bacon-Shone(1984)}]{Aitchison1984-tl}
Aitchison J, Bacon-Shone J (1984) Log contrast models for experiments with
  mixtures. Biometrika 71(2):323--330

\bibitem[{Argyriou et~al(2008)Argyriou, Evgeniou, and Pontil}]{Argyriou2008-cj}
Argyriou A, Evgeniou T, Pontil M (2008) Convex multi-task feature learning.
  Machine Learning 73(3):243--272

\bibitem[{Arumugam et~al(2011)Arumugam, Raes, Pelletier, Le~Paslier, Yamada,
  Mende, Fernandes, Tap, Bruls, Batto, Bertalan, Borruel, Casellas, Fernandez,
  Gautier, Hansen, Hattori, Hayashi, Kleerebezem, Kurokawa, Leclerc, Levenez,
  Manichanh, Nielsen, Nielsen, Pons, Poulain, Qin, Sicheritz-Ponten, Tims,
  Torrents, Ugarte, Zoetendal, Wang, Guarner, Pedersen, de~Vos, Brunak, Doré,
  {MetaHIT Consortium}, Antolín, Artiguenave, Blottiere, Almeida, Brechot,
  Cara, Chervaux, Cultrone, Delorme, Denariaz, Dervyn, Foerstner, Friss, van~de
  Guchte, Guedon, Haimet, Huber, van Hylckama-Vlieg, Jamet, Juste, Kaci, Knol,
  Lakhdari, Layec, Le~Roux, Maguin, Mérieux, Melo~Minardi, M'rini, Muller,
  Oozeer, Parkhill, Renault, Rescigno, Sanchez, Sunagawa, Torrejon, Turner,
  Vandemeulebrouck, Varela, Winogradsky, Zeller, Weissenbach, Ehrlich, and
  Bork}]{Arumugam2011-si}
Arumugam M, Raes J, Pelletier E, et~al (2011) Enterotypes of the human gut
  microbiome. Nature 473(7346):174--180

\bibitem[{Bien et~al(2021)Bien, Yan, Simpson, and Müller}]{Bien2021-cl}
Bien J, Yan X, Simpson L, et~al (2021) Tree-aggregated predictive modeling of
  microbiome data. Scientific Reports 11(1):14,505

\bibitem[{Boyd et~al(2011)Boyd, Parikh, and Chu}]{Boyd2011-ul}
Boyd S, Parikh N, Chu E (2011) Distributed optimization and statistical
  learning via the alternating direction method of multipliers. Foundations and
  Trends\textregistered \ in Machine Learning 3(1):1--122

\bibitem[{Chaudhury and Ramakrishnan(2015)}]{Chaudhury2015-wp}
Chaudhury KN, Ramakrishnan KR (2015) A new {ADMM} algorithm for the euclidean
  median and its application to robust patch regression. In: 2015 IEEE
  International Conference on Acoustics, Speech and Signal Processing (ICASSP),
  pp 1603--1607

\bibitem[{Combettes and Müller(2021)}]{Combettes2021-br}
Combettes PL, Müller CL (2021) Regression models for compositional data:
  General {Log-Contrast} formulations, proximal optimization, and microbiome
  data applications. Statistics in Biosciences 13(2):217--242

\bibitem[{Cowie et~al(1997)Cowie, Mosterd, Wood, Deckers, Poole-Wilson, Sutton,
  and Grobbee}]{Cowie1997-ib}
Cowie MR, Mosterd A, Wood DA, et~al (1997) The epidemiology of heart failure.
  European Heart Journal 18(2):208--225

\bibitem[{Cuevas-Sierra et~al(2020)Cuevas-Sierra, Riezu-Boj, Guruceaga,
  Milagro, and Martínez}]{Cuevas-Sierra2020-wy}
Cuevas-Sierra A, Riezu-Boj JI, Guruceaga E, et~al (2020) {Sex-Specific}
  associations between gut prevotellaceae and host genetics on adiposity.
  Microorganisms 8(6):938

\bibitem[{Dillon et~al(2016)Dillon, Frank, and Wilson}]{Dillon2016-ph}
Dillon SM, Frank DN, Wilson CC (2016) The gut microbiome and {HIV-1}
  pathogenesis: a two-way street. AIDS 30(18):2737--2751

\bibitem[{Gower(1971)}]{Gower1971-zp}
Gower JC (1971) A general coefficient of similarity and some of its properties.
  Biometrics 27(4):857--871

\bibitem[{Greenacre(2018)}]{Greenacre2018-dn}
Greenacre M (2018) Compositional Data Analysis in Practice. CRC Press

\bibitem[{Hallac et~al(2015)Hallac, Leskovec, and Boyd}]{Hallac2015-ss}
Hallac D, Leskovec J, Boyd S (2015) Network lasso: Clustering and optimization
  in large graphs. KDD: proceedings / International Conference on Knowledge
  Discovery \& Data Mining International Conference on Knowledge Discovery \&
  Data Mining 2015:387--396

\bibitem[{Haro et~al(2016)Haro, Rangel-Zúñiga, Alcalá-Díaz, Gómez-Delgado,
  Pérez-Martínez, Delgado-Lista, Quintana-Navarro, Landa, Navas-Cortés,
  Tena-Sempere, Clemente, López-Miranda, Pérez-Jiménez, and
  Camargo}]{Haro2016-xj}
Haro C, Rangel-Zúñiga OA, Alcalá-Díaz JF, et~al (2016) Intestinal
  microbiota is influenced by gender and body mass index. PLOS ONE
  11(5):e0154,090

\bibitem[{Kong et~al(2014)Kong, Fujimaki, Liu, Nie, and Ding}]{Kong2014-jm}
Kong D, Fujimaki R, Liu J, et~al (2014) Exclusive feature learning on arbitrary
  structures via $\ell_{1, 2}$-norm. In: Advances in neural information
  processing systems, pp 1655--1663

\bibitem[{Lengerich et~al(2018)Lengerich, Aragam, and Xing}]{Lengerich2018-zo}
Lengerich BJ, Aragam B, Xing EP (2018) Personalized regression enables
  sample-specific pan-cancer analysis. Bioinformatics 34(13):i178--i186

\bibitem[{Lin et~al(2014)Lin, Shi, Feng, and Li}]{Lin2014-ii}
Lin W, Shi P, Feng R, et~al (2014) Variable selection in regression with
  compositional covariates. Biometrika 101(4):785--797

\bibitem[{McMurdie and Holmes(2013)}]{McMurdie2013-do}
McMurdie PJ, Holmes S (2013) phyloseq: an {R} package for reproducible
  interactive analysis and graphics of microbiome census data. PLOS ONE
  8(4):e61,217

\bibitem[{Parikh and Boyd(2014)}]{Parikh2014-hd}
Parikh N, Boyd S (2014) Proximal algorithms. Foundations and Trends in
  Optimization 1(3):127--239

\bibitem[{Saraswati and Sitaraman(2015)}]{Saraswati2015-cg}
Saraswati S, Sitaraman R (2015) Aging and the human gut microbiota—from
  correlation to causality. Frontiers in microbiology 5:764

\bibitem[{Shi et~al(2016)Shi, Zhang, and Li}]{Shi2016-ys}
Shi P, Zhang A, Li H (2016) Regression analysis for microbiome compositional
  data. The Annals of Applied Statistics 10(2):1019--1040

\bibitem[{Tibshirani(1996)}]{Tibshirani1996-ak}
Tibshirani R (1996) Regression shrinkage and selection via the lasso. Journal
  of the Royal Statistical Society Series B 58(1):267--288

\bibitem[{Wang and Zhao(2017)}]{Wang2017-sy}
Wang T, Zhao H (2017) Structured subcomposition selection in regression and its
  application to microbiome data analysis. The Annals of Applied Statistics
  11(2):771--791

\bibitem[{Wu et~al(2011)Wu, Chen, Hoffmann, Bittinger, Chen, Keilbaugh, Bewtra,
  Knights, Walters, Knight, Sinha, Gilroy, Gupta, Baldassano, Nessel, Li,
  Bushman, and Lewis}]{Wu2011-os}
Wu GD, Chen J, Hoffmann C, et~al (2011) Linking {Long-Term} dietary patterns
  with gut microbial enterotypes. Science 334(6052):105--108

\bibitem[{Xu et~al(2015)Xu, Zhou, and Tan}]{Xu2015-yb}
Xu J, Zhou J, Tan PN (2015) {FORMULA}: {FactORized} {MUlti-task} {LeArning} for
  task discovery in personalized medical models. In: Proceedings of the 2015
  {SIAM} International Conference on Data Mining ({SDM}). Proceedings, Society
  for Industrial and Applied Mathematics, p 496--504

\bibitem[{Yamada et~al(2017)Yamada, Koh, Iwata, Shawe-Taylor, and
  Kaski}]{Yamada2017-cg}
Yamada M, Koh T, Iwata T, et~al (2017) {Localized Lasso for {High-Dimensional}
  Regression}. In: Singh A, Zhu J (eds) Proceedings of the 20th International
  Conference on Artificial Intelligence and Statistics, Proceedings of Machine
  Learning Research, vol~54. PMLR, pp 325--333

\bibitem[{Zeng et~al(2019)Zeng, Li, He, Li, Yang, Zhao, Liu, Wang, Sun, Feng,
  Wang, Chen, Zheng, Yang, Sun, Xu, Wang, Kenney, Jiang, Gu, Li, Zhou, Li, and
  Dai}]{Zeng2019-nx}
Zeng Q, Li D, He Y, et~al (2019) Discrepant gut microbiota markers for the
  classification of obesity-related metabolic abnormalities. Scientific Reports
  9(1):13,424

\end{thebibliography}


\end{document}